\documentclass{raa}

\usepackage{times}
\usepackage{graphicx}
\usepackage{natbib}
\usepackage{amssymb,amsmath}      
\usepackage{textcomp}
\usepackage{multirow}
\usepackage{mathtools}
\bibpunct{(}{)}{;}{a}{}{,}

\usepackage[a4paper=true,dvipdfm=true,pagebackref=true]{hyperref}
\hypersetup{pdftitle = The title of my PDF, pdfauthor = My name, pdfsubject= The subject, pdfkeywords = keyword1 keyword2 keyword3}
\hypersetup{colorlinks = true, linkcolor = green, anchorcolor = red, citecolor = blue, filecolor = red, urlcolor = red}

\begin{document}

\title{NUV Star Catalogue from the Lunar-based Ultraviolet Telescope Survey. First Release}

\volnopage{ {\bf 2016} Vol.\ {\bf X} No. {\bf XX}, 000--000}
\setcounter{page}{1}

\author{Xian-Min Meng, Xu-Hui Han, Jian-Yan Wei, Jing Wang, Li Cao, Yu-Lei Qiu, Chao Wu, Jin-Song Deng, Hong-Bo Cai, Li-Ping Xin}

\institute{Key Laboratory of Space Astronomy and Technology, National Astronomical Observatories, Chinese Academy of Sciences, Beijing 100012, China; 
{\it mengxm@nao.cas.cn; hxh@nao.cas.cn}\\
\vs \no
{\small Received April 24, 2016; accepted June 29, 2016}
}

\abstract{We present a star catalogue extracted from the 
Lunar-based Ultraviolet Telescope (LUT) survey program. 
LUT's observable sky area is a circular belt around the Moon's north pole,
and the survey program covers a preferred area for about 2400\,$\deg^2$ 
which includes a region of the Galactic plane.
The data is processed with an automatic pipeline
which copes with stray light contamination, artificial sources, cosmic rays, 
flat field calibration, photometry and so on. 
In the first release version, the catalogue provides high confidence sources 
which have been cross-identified with Tycho-2 catalogue.
All the sources have signal-to-noise ratio larger than 5,
and the corresponding magnitude limit is typically 14.4\,mag,
which can be deeper as $\sim$16\,mag if the stray light contamination is in the lowest level.
A total number of 86,467 stars are recorded in the catalogue.
The full catalogue in electronic form is available on line.
}

\authorrunning{X.-M. Meng et al. } 
\titlerunning{Star Catalogue from LUT Survey. I}
\maketitle

\section{Introduction}
\label{sec:Intro}

Lunar-based Ultraviolet Telescope (LUT) is the first robotic telescope 
deployed on the moon's surface, 
and is loaded inside the lander of China's Chang'e-3 lunar exploration program \citep{Ip2014RAA}.
LUT and the lander are located at 44.12$\degr$N and 19.52$\degr$W 
on a basin of the moon named Sea Of Showers.
LUT is an imaging telescope working at a characteristic near-ultraviolet (NUV) band.
Since the successful launch and landing of Chang'e-3 in December 2013,
LUT had finished the task of its one-year's mission phase and continued to work stably
for another one year.
The stability of its performance has been verified by 
an 18-month magnitude zero point (zp) calibration work. 
The photometric calibration gives $zp=17.53\pm0.05$\,mag, 
which is highly consistent with the results of 
the first 6-months of $zp=17.52\pm0.07$\,mag \citep{WangJ2015zp}.

One of LUT's main scientific objectives is to perform a sky survey for an area 
about 2400\,$\deg^2$ \citep{Caoli2011}.
Tens of star catalogues in NUV bands have already been published, which are contributed by 
Galaxy Evolution Explorer (GALEX), Hubble Space Telescope (HST), etc.
GALEX has an all sky survey project named All-Sky Imaging survey (AIS), 
whose detection limit is $\sim$21\,mag at an NUV band.
GALEX avoids the Galactic plane during the prime mission phase because of its high-countrate safety limits.
Its latest survey covers regions in the Galactic plane, 
but these data are not reachable yet in the public archive \citep{BianchiGALEX2014}.
LUT survey covers a part of the low Galactic latitude region within its available sky area,
so it would be helpful for future researches for this region.
Further more, the survey observation strategy of LUT enables the telescope 
to revisit some sky areas for more than 10 times,
so it is potentially possible to find variable stars through further data mining.

The LUT survey data are processed with an automatic pipeline,
which is inherited from LUT's pointing observation data processing pipeline 
\citep[see][]{Meng2015ApSS}. 
The main different parts from the pointing program are: 
(1) the survey data processing has to clean off all the cosmic rays,
so it has to perform image stacking; 
(2) a series of processes have to be carried out to clean off artificial sources 
that have arisen from stray light residuals.

The first release of LUT survey data product, as described in this paper,
is a star catalogue covering the whole LUT's available sky area, 
and is cross-identified with an optical star catalogue Tycho-2 \citep{Tycho2000}.
Tycho-2 is the reference catalogue for LUT's astrometry, 
and its positional and photometric data are very helpful for LUT to remove artificial sources.

After a quick look of the LUT instrument in Sect.~\ref{sec:Inst},
the survey observation strategy and its footprints are described in Sect.~\ref{sec:obs}.
The details of the survey data processing pipeline are described in Sect.~\ref{sec:Pipeline}, 
and the performances in various aspects are also shown.
Compilation of the catalogue and some statistical results 
are presented in Sect.~\ref{sec:catalogue}.
Discussion on Galactic extinction and aperture correction is given in Sect.~\ref{sec:Discuss}.

\section{LUT Instrument}
\label{sec:Inst}

\begin{figure}[!htbp]
  \centering
  \includegraphics[width=0.8\columnwidth]{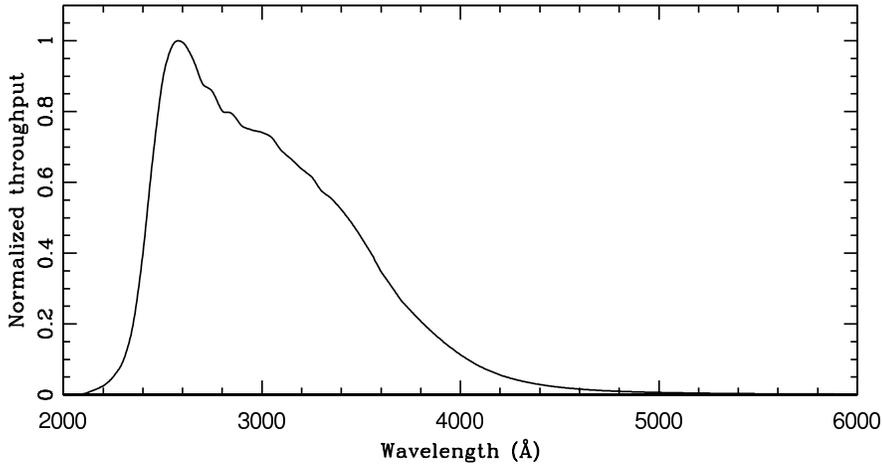}
  \caption{Normalized throughput curve of LUT system \citep{Wangjing2015cali}.}
  \label{fig:through}
\end{figure}

LUT is a Ritchey-Chretien telescope working at a Nasmyth focus.
A flat mirror is mounted on a two-dimensional gimbal in front of the telescope aperture
for pointing and tracking \citep[see][Fig.~1]{Wangjing2015cali}.
A UV-enhanced back-illuminated AIMO CCD is mounted on the fucal plane,
and UV coating is applied on one lense of the field corrector as the UV filter.
The final throughput of the LUT system has a peak value of $\sim$8\% at 2500\,\AA.
The throughput curve is shown in Figure~\ref{fig:through}.
Two pairs of LEDs are installed crosswise (one as backup) on the front inside wall of the camera, 
which is used to illuminate the CCD through a ring-like diffusion glass for flat field calibration.
The LEDs emit at spectral wavelength 286\,nm and the spectral width is $\sim$12\,nm.
Further details of the instrumentation and system performance 
are described by \cite{Caoli2011} and \cite{Wangjing2015cali}.
The basic parameters of the LUT instruments are listed in Table~\ref{tab:inst}.

\begin{table}
\centering
\caption[]{Basic parameters of LUT instruments.}
\label{tab:inst}
\setlength{\tabcolsep}{1pt}
 \begin{tabular}{l@{\extracolsep{2em}}l}
  \hline\noalign{\smallskip}
  Item & Value\\
  \hline\noalign{\smallskip}
  Telescope aperture (mm)                      & 150\\
  Focal ratio                                  & 3.75\\
  Field of view                                & 1.35\textdegree$\times$1.35\textdegree\\
  CCD size (imaging area; pixel)               & 1024$\times$1024\\
  CCD pixel scale (arcsec/pixel)               & 4.76\\
  Readout noise ({\it e}$^{\rm -1}$/pixel)     & 8\\
  Dark current ({\it e}$^{\rm -1}$/pixel/s)    & $<$0.2\\
  Passband (nm)                                & 245--345\\
  Azimuth range of mirror rotation             & \lbrack-28\textdegree, +13\textdegree\rbrack\\
  Altitude range of mirror rotation            & \lbrack+20\textdegree, +38\textdegree\rbrack\\
  \noalign{\smallskip}\hline
\end{tabular}
\end{table}

\section{Survey Observations}
\label{sec:obs}

LUT takes advantage of its own flat mirror and the moon's self-rotation 
to change the survey area of the sky, 
and hence has surveyed a wide circular belt around the moon's north pole.
This area is generally between latitude 60$\degr$N and 80$\degr$N 
in the Moon-fixed coordinate system\footnote{The origin is the barycenter of the Moon; 
the {\it z}-axis coincides with the mean rotational axis (pole) of the Moon; 
the zero latitude circle is the equatorial plane;
the prime meridian (0$\degr$\ longitude) is defined by the mean Earth direction \citep{Qi2015}.},
and the total survey area is $\sim$2400$\deg^2$.
During each Lunar daytime, LUT's survey program may have 13 observation sequences
which always contains 36 different pointings.
The pointing directions are always shifted by 1$\degr$ and overlap with each other,
which is designed for the completeness of sky coverage.
Each sky area was observed for 10 minutes with single exposure time 30\,s
and idle for 30\,s (designed to meet the requirement of the total data size limit of Chang'e-3).
During the observation for each sky area (10 minutes), 
the direction of the pointing keeps stationary relative to the Moon's surface.
In each lunar month the observation time of the survey program adds up to 72 hours.
The observation time sequences of the pointings are calculated 
at each beginning of lunar daytime, 
and also a survey mosaic is simulated by a separately developed code.

\begin{figure}[!htbp]
 \centering
 \includegraphics[width=0.8\textwidth]{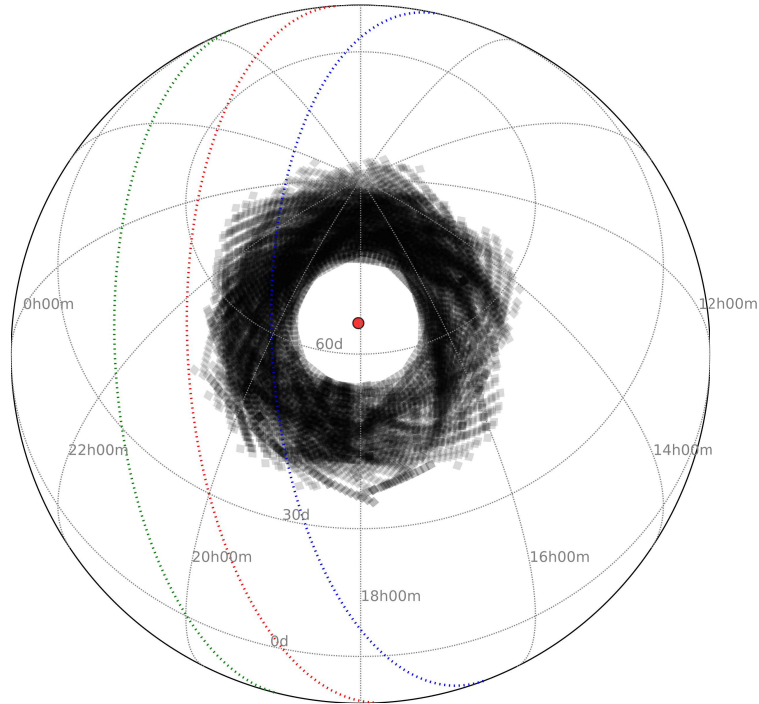}
 \caption{The LUT survey footprints, which covering $\sim$2400\,$\deg^2$. 
 The red filled circle marks the position of lunar north pole in the epoch of Jan.\ 1\ 2015; 
 the black filled quadrangles marks individual sky coverage of LUT's FOV in different time;
 the red dotted line marks the Galactic plane as $b=0\degr$;
 the blue and green dotted line marks the $b=+15\degr$ and $b=-15\degr$ Galactic 
 latitudes, respectively.}
 \label{fig:footprint}
\end{figure}

The sky coverage of the LUT survey is illustrated in Figure~\ref{fig:footprint}. 
The red filled circle marks the position of the lunar north pole in the epoch of Jan. 1 2015 
\footnote{Calculated through The IAU Working Group on Cartographic Coordinates and Rotational Elements (WGCCRE) report \citep{wgccre2011}.};
the black filled quadrangles mark individual sky coverage of LUT's field of view (FOV) 
in J2000 coordinates in different time;
the red dotted line marks the Galactic plane as $b=0\degr$;
the blue and green dotted lines mark the $b=+15\degr$ and $b=-15\degr$ Galactic 
latitudes, respectively.
Figure~\ref{fig:footprint} shows that a part of the low Galactic latitude ($b<15\degr$) region is 
covered by LUT survey, which provides valuable measurement of stars in this region at the NUV band.

\section{Data Processing For Survey Program}
\label{sec:Pipeline}

The survey data processing pipeline is an automatic system.
It is inherited from the pipeline of LUT's pointing observation programs \citep{Meng2015ApSS},
so the main strategy, algorithms, and many procedures are similar to the latter.
It starts from each raw image, and ends to produce a star catalogue of each specified sky area.
The outline of the pipeline is listed in Table~\ref{Tab:Pipeline}.
In this section, 
the procedures that are similar to the pointing programs will be shortly covered firstly,
and the extra parts of the survey pipeline will be described in detail in the following subsections.

\begin{table}[!htbp]
\centering
\caption[]{Outline Of Survey Data Processing Pipeline To Produce Level-2 Data}
\begin{tabular}{|l|}
\hline
{\bf Outline of Survey Data Processing Pipeline}\\
\hline
0. Data preparation\\
1. Overscan correction \& imaging area trimming\\
{\it Stray Light Removing}\\
~~~~2. Image grouping according to pointing direction and time period\\
~~~~3. Image combination to produce stray light pattern template\\
~~~~4. Each image subtract its stray light template\\
5. Flat fielding making use of dithering observation\\
6. Image stacking\\
7. Astrometry\\
8. Source Extraction\\
{\it Profile Measurement and Source Cleaning (1st Cleaning)}\\
~~~~9. Remove abnormal objects which are on edges of images, have negative fluxes, ELONGATION$>$3, etc.\\
~~~~10. Measure FWHMs of Moffat profile and keep objects which have 1.3$<$MFWHM$<$4\\
~~~~11. Determine typical MFWHM for an image group and assign to FWHM$_{\rm med}$\\
{\it Photometry}\\
~~~~12. Aperture photometry\\
~~~~13. Artificial sources cleaning (2nd cleaning) through clustering analysis method\\
~~~~14. PSF photometry\\
15. Catalogues archiving\\
\hline
\end{tabular}
\label{Tab:Pipeline}
\end{table}

The LUT observation suffers from stray light contamination 
caused by sunlight being scattered by the cabin and the telescope.
The pattern of stray light can be removed by subtracting a stray light template
generated from the observation images.
After been grouped according to their pointing directions and observation times,
the images are combined with ``median'' algorithm to produce stray light templates. 
Because the flat mirror keeps stationary during each pointing period
and the Moon's self-rotation leads to little offsets ($\sim$4\,pixel) of the source positions
between adjacent exposures,
image combination can produce stray light templates without any point sources.
Then, stray light can be removed by subtracting the template from the original images.
Bias and dark current are also subtracted along with the stray light pattern.

Flat field images are produced by making use of dithering observations.
We use the internal LED lamps to produce flat images which will be flattened to correct 
pixel-to-pixel relative nonuniformity, and employ dithering observation of 
individual standard stars to sample the large-scale nonuniformity in the field of view.
So the flat fielding correction is accomplished in two steps, a pixel-to-pixel flat fielding,
and a large-scale flat fielding using an image produced by 2-D surface fitting 
to the sampling fluxes of the calibration star.

After image stacking (Sect.~\ref{subsec:combine}) 
and astrometry calibration (Sect.~\ref{subsec:astrometry}),
source candidates are extracted from each stacked image
by {\sc SExtractor} \citep{Bertin1996A&AS}.
Some criteria are adopted to exclude abnormal 
or irrelative source candidates for subsequent procedures.
The first step is to clean the candidates which 
(1) are in the margin regions of the images;
(2) have negative fluxes measured by {\sc SExtractor};
(3) have ELONGATION$>3$ (major-axis/minor-axis$>$3);
(4) have photometry failure flags;
(5) have abnormal background values, and
(6) have FWHMs$<$1.3\,pix (full widths at half maximum$<$1.3\,pix) or FWHMs$>$4\,pix, 
according to the distribution of FWHMs 
which peaks at 2\,pix and extends to 3.5\,pix.
The other step is performed after aperture photometry, and its purpose is 
to remove artificial sources arisen from stray light residual (see Sect.~\ref{subsec:fakeclean}).

A typical value of FWHM is determined for a group of images
as the unit of aperture radius of photometry. 
The typical FWHM of each group is denoted as FWHM$_{\rm med}$.
The determining method of FWHM$_{\rm med}$ is described by \cite{Meng2015ApSS}.
When aperture photometry and PSF photometry (see Sect.~\ref{subsec:phot}) are finished, 
a star catalogue recording all the output results is generated for every image. 
These catalogues are merged to generate the preliminary source catalogue of LUT.

The data processing procedures that are different from the pointing program 
are described in the following subsections.

\subsection{Image Stacking}
\label{subsec:combine}
The image stacking procedure combines the images by computing the ``median'' value of every pixels.
The main purposes of the image stacking procedure are: 
(1) to clean off cosmic rays; (2) to increase detection depth and signal-to-noise ratio (SNR); 
(3) to solve the charge transfer efficiency (CTE) problem.
The CTE problem is found on the LUT CCD where residual charges may left after the readout.
It may happen after some types of cosmic ray strikes, and disappear after one more exposure.
Therefore, to filter out cosmic rays and their residual charges,
the number of coadded images shoul be not less than 5,
and typically, the number is set to be 8,
for it is the typical total number of images for each survey patch observation.

Before performing ``median'' combination, 
the images are grouped according to the pointing directions, 
and the time duration of a group is restricted to be within 10 minutes.
Then, the images of each group are aligned according to the physical positions 
of the point sources on the images. Finally, the ``median'' combination is performed.

\subsection{Astrometry}
\label{subsec:astrometry}

The astrometry procedure for the survey images follow the same procedure in the pointing program
\citep{Meng2015ApSS}. In short, after converting the epoch of Tycho-2 catalogue to the current
and revising the positions of Tycho-2 stars for proper motion, 
the geometrical distribution of 5--10 bright stars on an LUT image are matched with
the Tycho-2 catalogue. 
In addition to cross matching their positions, their brightnesses are used
as another cross-matching criterion to exclude false matching which correlates two irrelevant stars.
International Celestial Reference System (ICRS) at epoch of J2000 is adopted 
for the astrometry result whose precision is typically about 1\arcsec 
(for the stars that are used in the astrometry).

\subsection{Photometry}
\label{subsec:phot}

LUT photometries adopt AB magnitude system defined by \citet{Oke1983},
and the magnitudes are obtained as
\begin{equation}
 m_{\rm LUT} = m_{\rm 0,LUT} - 2.5 \log f_{\rm LUT}
\end{equation}
where $m_{\rm 0,LUT}$ is the magnitude zero point of LUT,
$f_{\rm LUT}$ is the flux density in unit of ergs/s/cm$^2$/Hz.
The zero point of LUT is determined from the observations of the standard stars provided by 
International Ultraviolet Explorer (IUE) mission \citep{WuCC1998IUE}.
The theoretical spectra of these standards were extracted from the {\sc ATLAS9} 
stellar atmosphere models \citep{Kurucz2003} 
and their absolute fluxes were determined according to their {\it V}-band magnitudes 
recorded in the SIMBAD database.
After constructing the theoretical spectra for the standards, 
the expected magnitude of star {\it i}, $m_{\rm i,exp}$ in LUT system, 
was calculated by involving the pre-determined throughput curve of LUT.
With an arbitrary magnitude zero point $m_{\rm 0,inst}$, 
the instrumental magnitude of star {\it i} was measured to be $m_{\rm i,inst}$, 
and so the zero point of LUT can be calculated through 
$m_{\rm 0,LUT}=m_{\rm 0,inst}-(m_{\rm i,inst}-m_{\rm i,exp})$.
The statistic result of all the calibration stars observed at different times gives 
the final magnitude zero point of LUT as $m_{\rm 0,LUT} = 17.52\pm0.05$, 
and this value remains stable in an 18-month operational performance test.
For details of the LUT photometry calibration and 18-months' stability of the zero point please refer to
\citet{Wangjing2015cali} and \citet{WangJ2015zp}. 

Aperture photometries are performed with an aperture radius of 2$\times{\rm FWHM_{med}}$,
where ${\rm FWHM_{med}}$ is denoted as the stars' typical FWHM.
The background annulus for aperture photometry adopt 6$\times{\rm FWHM_{med}}$ 
and 9$\times{\rm FWHM_{med}}$ as inner and outer radii, respectively.
Aperture photometries are performed by {\sc PyRAF.apphot.phot} code.
SNR is calculated as
\begin{equation}
 {\rm SNR} = \dfrac{F}{\sqrt{\dfrac{F}{G} + A \times \sigma^2 + \dfrac{A^2\times\sigma^2}{N_{\rm sky}}}}
\end{equation} 
where $F$ is the total number of counts excluding the background in the aperture, 
$G$ is the gain of CCD (electrons per ADU), 
$A$ is the area in aperture, 
$\sigma$ is the standard deviation of the background, 
$N_{\rm sky}$ is the pixel count of background. 
The uncertainty of aperture photometry is calculated as 
${\rm m_{err,LUT}} = 1.0857/{\rm SNR}$ and is typically ${\rm m_{err,LUT}}\sim0.006$\,mag 
for a star of $m_{\rm LUT}=10$\,mag (30\,s exposure).
It rises to about 0.2\,mag at magnitude $\sim$14.4 which is the 5$\sigma$ detection.
The relation between photometric accuracies and brightness is shown in Figure~\ref{fig:statis}.

\begin{figure}[!htbp]
  \centering
  \includegraphics[width=0.8\columnwidth]{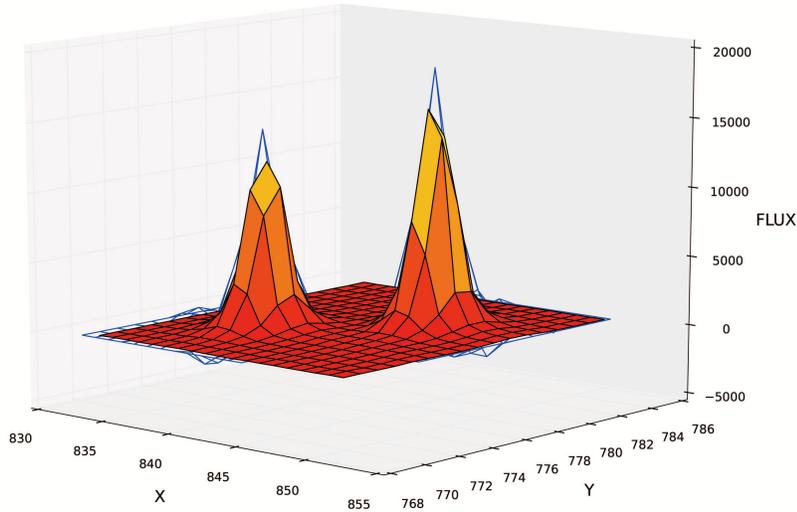}
  \caption{PSF modeling for a double star 16~Cyg.}
  \label{fig:psfphot}
\end{figure}

PSF photometries are also performed to provide more reliable measurements for crowded stars.
The pipeline has been tested to have the capability to extract positions of contacting 
and crowded stars and to model their PSF profiles.
The actual effect of the PSF photometry of a double star 16~Cyg
is illustrated in Figure~\ref{fig:psfphot}.

\subsection{Artificial Sources Cleaning}
\label{subsec:fakeclean}
\begin{figure}[!htbp]
  \centering
  \includegraphics[width=0.5\columnwidth]{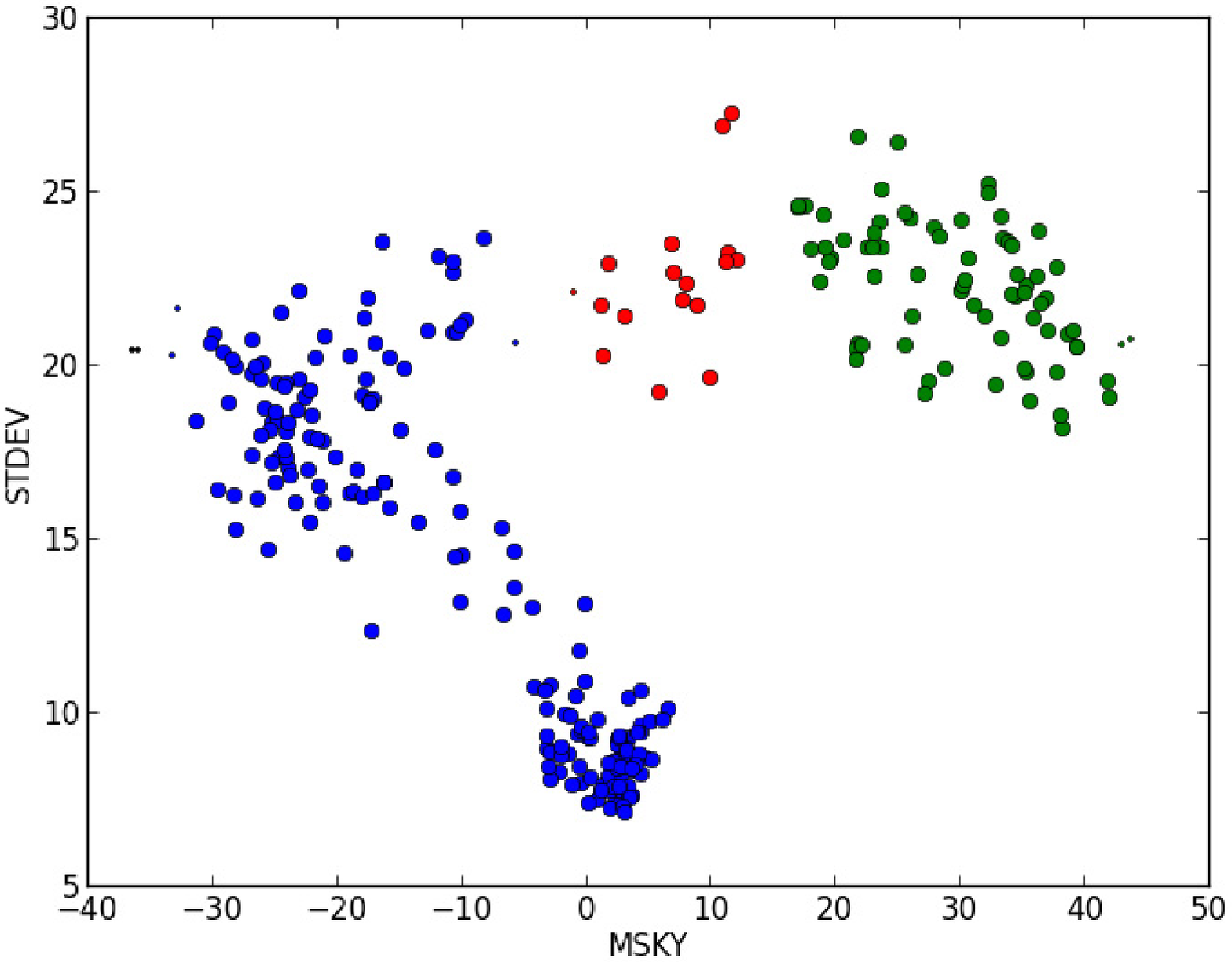}\\
  \includegraphics[width=0.5\columnwidth]{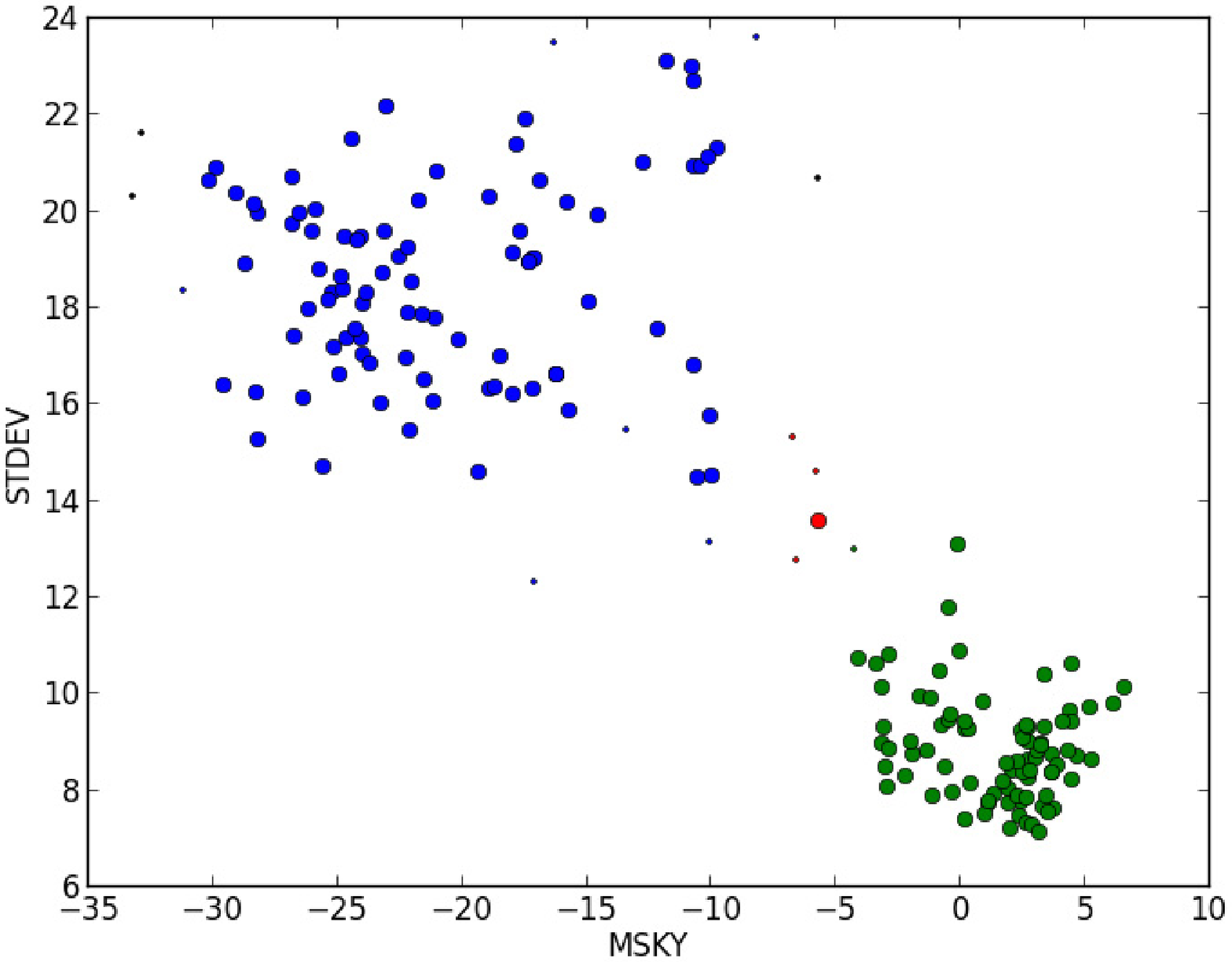}\\
  \includegraphics[width=0.5\columnwidth]{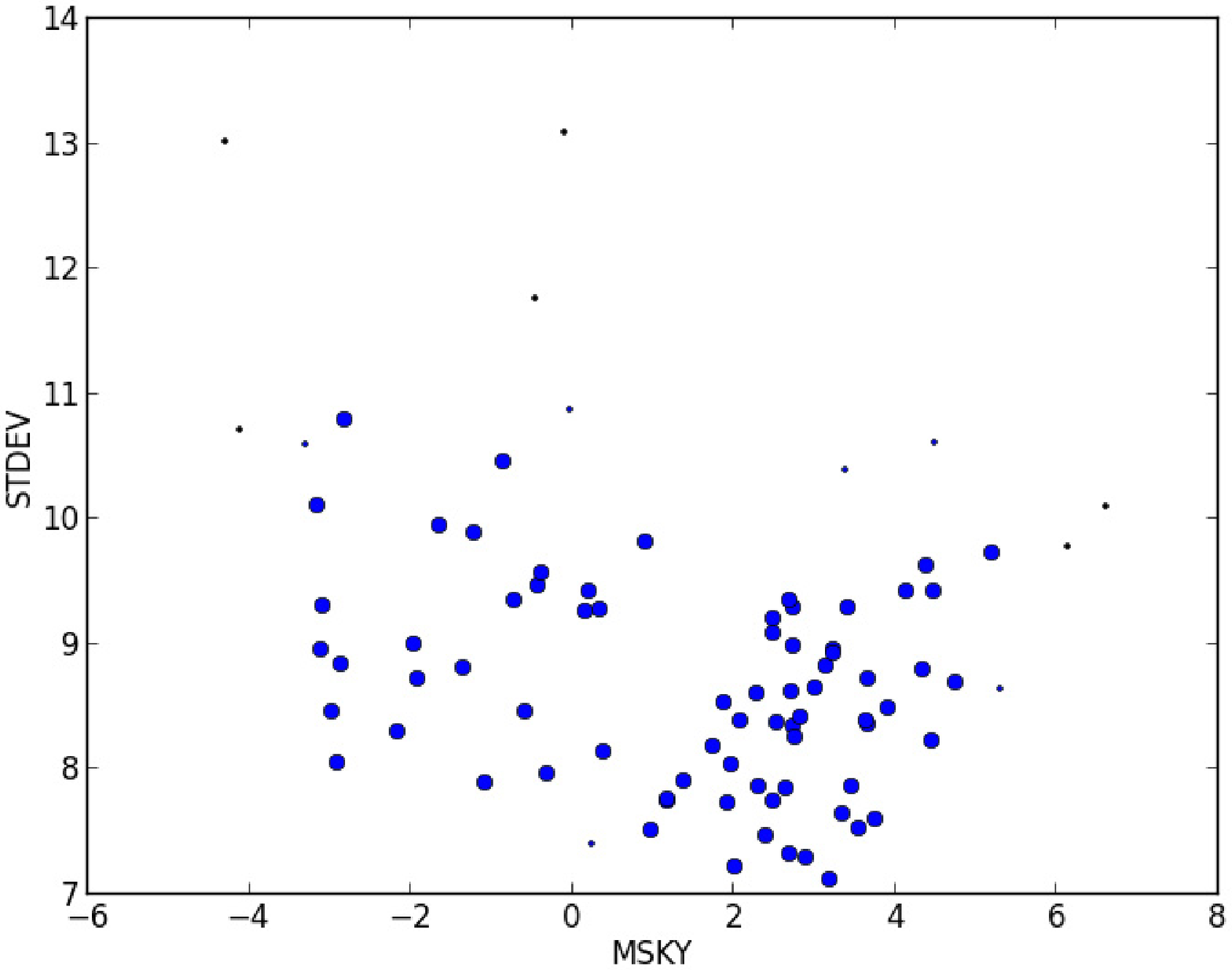}
  \caption{An example of the three-step iteration of clustering analysis in MSKY--STDEV space,
  where ``MSKY'' is the source's background value, and ``STDEV'' is the standard deviation of ``MSKY''.
  The colored and filled circles mark different clusters in each iteration, 
  smaller circles are non-core samples that are still part of a cluster,
  and the black points mark the outliers. 
  There is no inheritance for the colors among these panels.}
  \label{fig:clustering}
\end{figure}

\begin{figure}[!htbp]
  \centering
  \includegraphics[width=0.45\columnwidth]{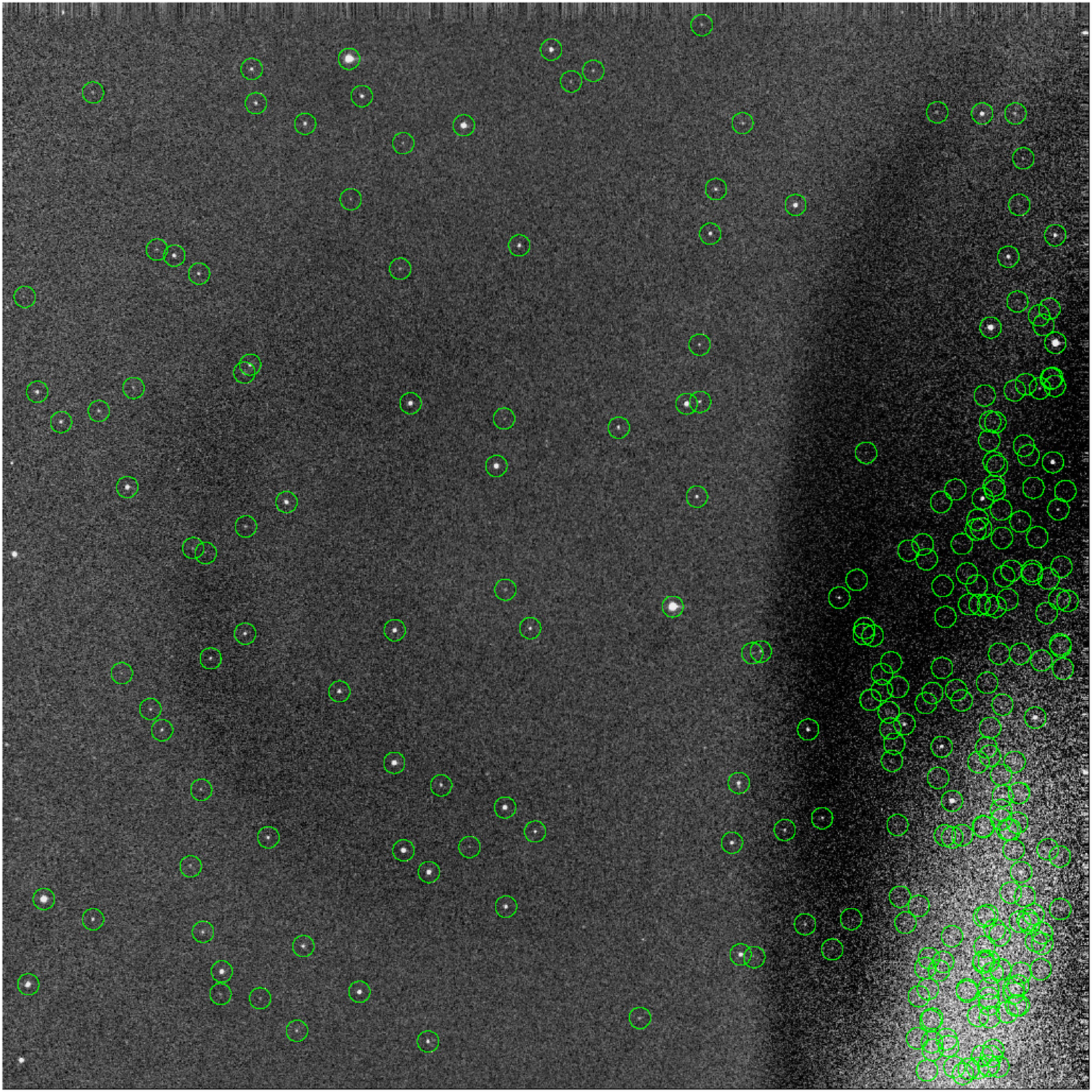}\hspace{10pt}
  \includegraphics[width=0.45\columnwidth]{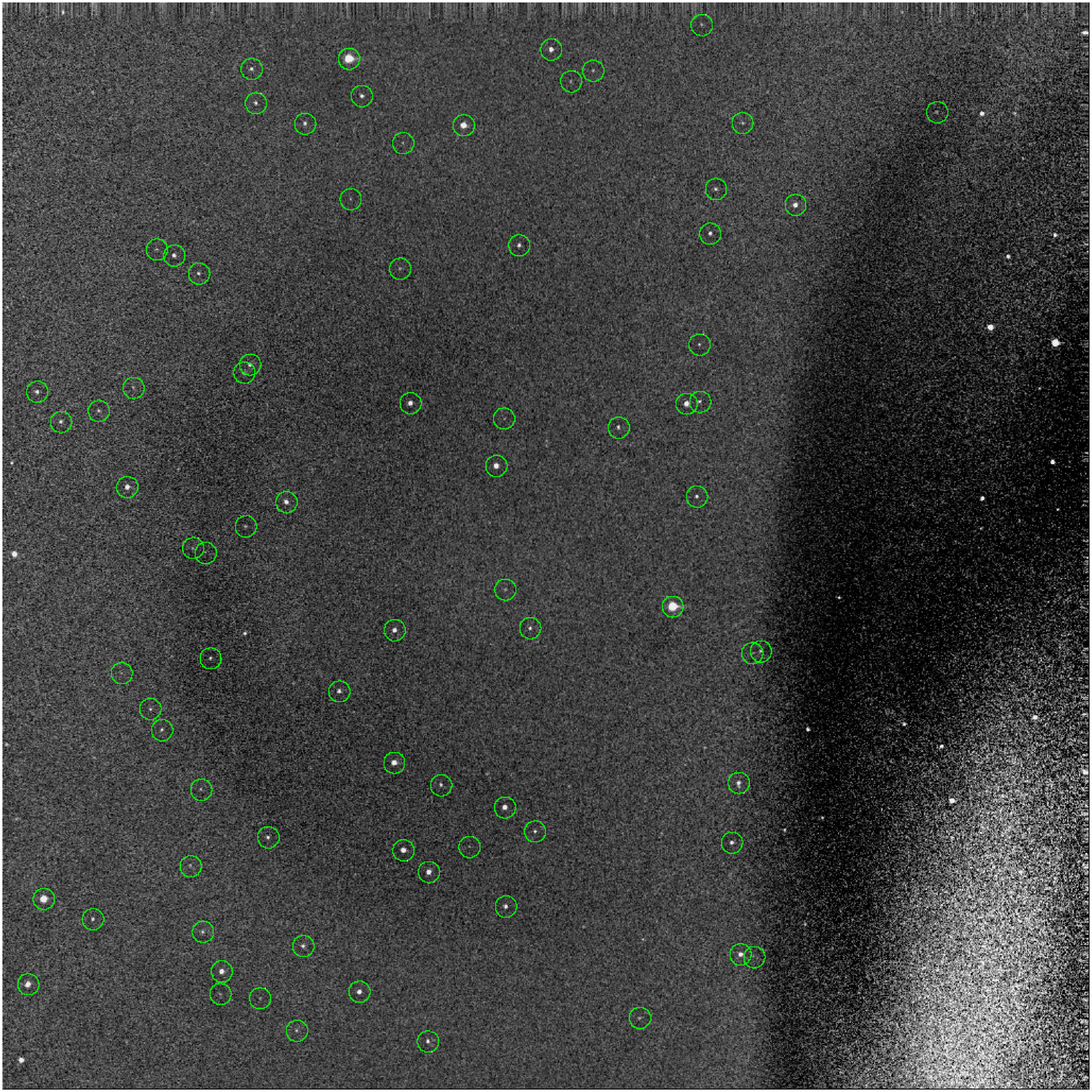}
  \caption{The effect of cleaning artificial sources.
  In the left panel, there are residuals of stray light removing on the right side
  and they lead to extracting artificial sources. After the artificial sources cleaning 
  procedures, most of them are filtered out.}
  \label{fig:clusterresult}
\end{figure}

The stray light removing procedure sometimes leaves residual pattern on the outcome images, 
if stray light varies with time and the current strategy can not remove stray light totally. 
Stray light residuals may cause source extraction to give false sources, 
which are actually noises of the residuals. 
To identify such false sources, we employ clustering analysis in a 2-D parameter space.
The sources' background values ``MSKY'' and the standard deviations ``STDEV'' of ``MSKY''
are used for clustering. 
The machine learning package {\sc scikit-learn} \citep{scikit-learn} and its {\sc dbscan} 
clustering method are used for this task.
Real celestial objects are selected 
as the cluster locating within a region which is the closest to (0,0) 
in the MSKY--STDEV space\footnote{``MSKY'' should be close to 0 because the background 
has been subtracted along with the stray light;
``STDEV'' is required to be the smallest where the noise of the background is the lowest.}. 
The scope of this region should be smaller than 5$\times$8 on an experimental basis.
The clustering analysis for each image is performed by no more than three iterations.
Figure~\ref{fig:clustering} shows an example of a three-step iteration,
and its result is illustrated in the right panel of Figure~\ref{fig:clusterresult}, 
which has cleaned off the artificial objects from the left panel.
Furthermore, an SNR cut is applied to select high confidence detections,
which have ${\rm SNR} \geq 5$, since the false sources always have lower SNR.

\section{The Catalogue}
\label{sec:catalogue}

In the first released version of LUT survey catalogue, we provide high confidence sources
with ${\rm SNR} \geq 5$, 
and sources that have been cross matched with the Tycho-2 catalogue \citep{Tycho2000}.
The catalogue with 86,467 lines is available as catalog I/335 from CDS\footnote{http://cdsarc.u-strasbg.fr/viz-bin/Cat?cat=I/335} (the Strasbourg astronomical Data Center).

\subsection{Catalogue Compilation}
\label{subsec:catgen}
The first step of catalogue construction is to deal with multiple observations of the sources,
which have been revisited by the survey for many times and have several records in the catalogue.
The representative record of each source is selected to be the first one in the primitive catalogue. 
This step generates $\sim$158,354 individual sources observed by LUT.

\begin{figure}[!htbp]
  \centering
  \includegraphics[width=0.6\columnwidth]{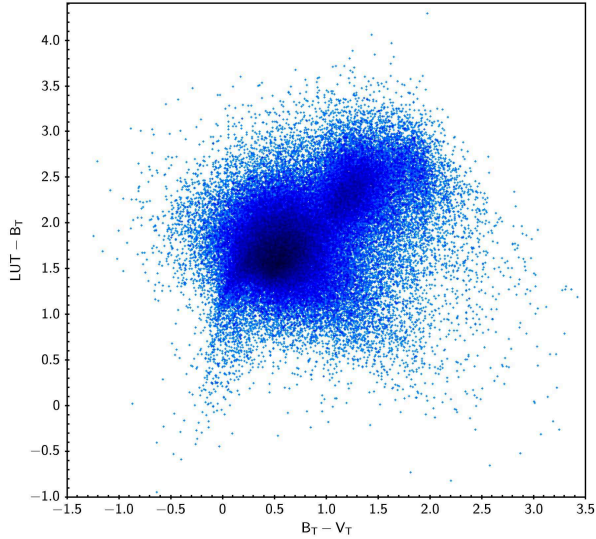}
  \caption{A color-color diagram of ${\rm LUT}-B_T$ vs. $B_T - V_T$ of the catalogue.}
  \label{fig:ccfig}
\end{figure}

The second step is to positionally cross-match the LUT catalogue with the Tycho-2 catalogue.
The purpose of this step is to identify the LUT stars in the established databases.
The cross matching is initially performed as finding the nearest counterparts 
between the two catalogues within 10\arcsec, i.e. nearly two pixels as separation distance.
The distribution of the separation distance shows that
more than 99\% resultant sources have been included within that of 4\arcsec, 
which is slightly less than the pixel scale of 4.76\arcsec\ of LUT.
So we adopt 4\arcsec\ as the criterion of positional matching.
This criterion results in a total number of 86,467 entries of the final catalogue.
Proper motions of the stars were not considered, because such factors between epoch J2000
and the current are negligible compared to the cross-matching distance of 4\arcsec.
The result of cross matching can be examined by a color-color diagram 
in Figure~\ref{fig:ccfig}. 
The dense region near $LUT-B_T\sim1.5$\,mag is typical for $B_T-V_T\sim0.5$\,mag stars
(F, G type), which is consistent with the theoretical estimation performed by 
Han et al. (2016, in preparation).
The number density drops around $LUT-B_T=2.0$\,mag, 
which is consistent with the 5$\sigma$ detections of LUT's 14.4\,mag 
and Tycho-2's $B_T\sim12.5$\,mag respectively. 

\subsection{Catalogue Format}
\label{subsec:catafmt}

The catalogue provides data from the LUT survey and the corresponding Tycho-2 records.
Table~\ref{tab:catafmt} describes the format of the catalogue.
\begin{table}[!htbp]
\centering
\caption[]{Catalogue format.}
\label{tab:catafmt}
\setlength{\tabcolsep}{1pt}
 \begin{tabular}{l@{\extracolsep{2em}}ll}
  \hline\noalign{\smallskip}
  Column Header & Units   & Description\\
  \hline\noalign{\smallskip}
  LUTID         & ---     & LUT object identifier\\
  RA\_LUT       & degree  & Right ascension in LUT records, ICRS, epoch=J2000\\
  DEC\_LUT      & degree  & Declination in LUT records, ICRS, epoch=J2000\\
  FWHM          & pix     & FWHM of the star, measured by {\sc PyRAF.psfmeasure} code\\
  R\_AP         & pix     & Aperture radius used for the aperture photometry\\
  MAG\_LUT      & mag     & Magnitude at LUT band measured with 2$\times$FWHM aperture radius\\
  MERR\_LUT     & mag     & Uncertainty of MAG\_LUT\\
  PSF\_LUT      & mag     & Magnitude measured through PSF modeling\\
  PSF\_ERR      & mag     & Uncertainty of PSF\_LUT\\
  JD            & day     & Julian date of LUT observation's UT time\\
  TYCHOID       & ---     & Tycho-2 identifier\\
  RA\_TYC       & degree  & Right ascension in Tycho-2 records, ICRS, epoch=J2000\\
  DEC\_TYC      & degree  & Declination in Tycho-2 records, ICRS, epoch=J2000\\
  BTMAG         & mag     & Tycho-2 $B_T$ magnitude\\
  VTMAG         & mag     & Tycho-2 $V_T$ magnitude\\
  SEPARATION    & arcsec  & Separation distance of the positions between LUT and Tycho-2\\
  \noalign{\smallskip}\hline
\end{tabular}
\end{table}

\subsection{Statistics}
\label{subsec:statis}

\begin{figure}[!htbp]
  \centering
  \includegraphics[width=0.6\columnwidth]{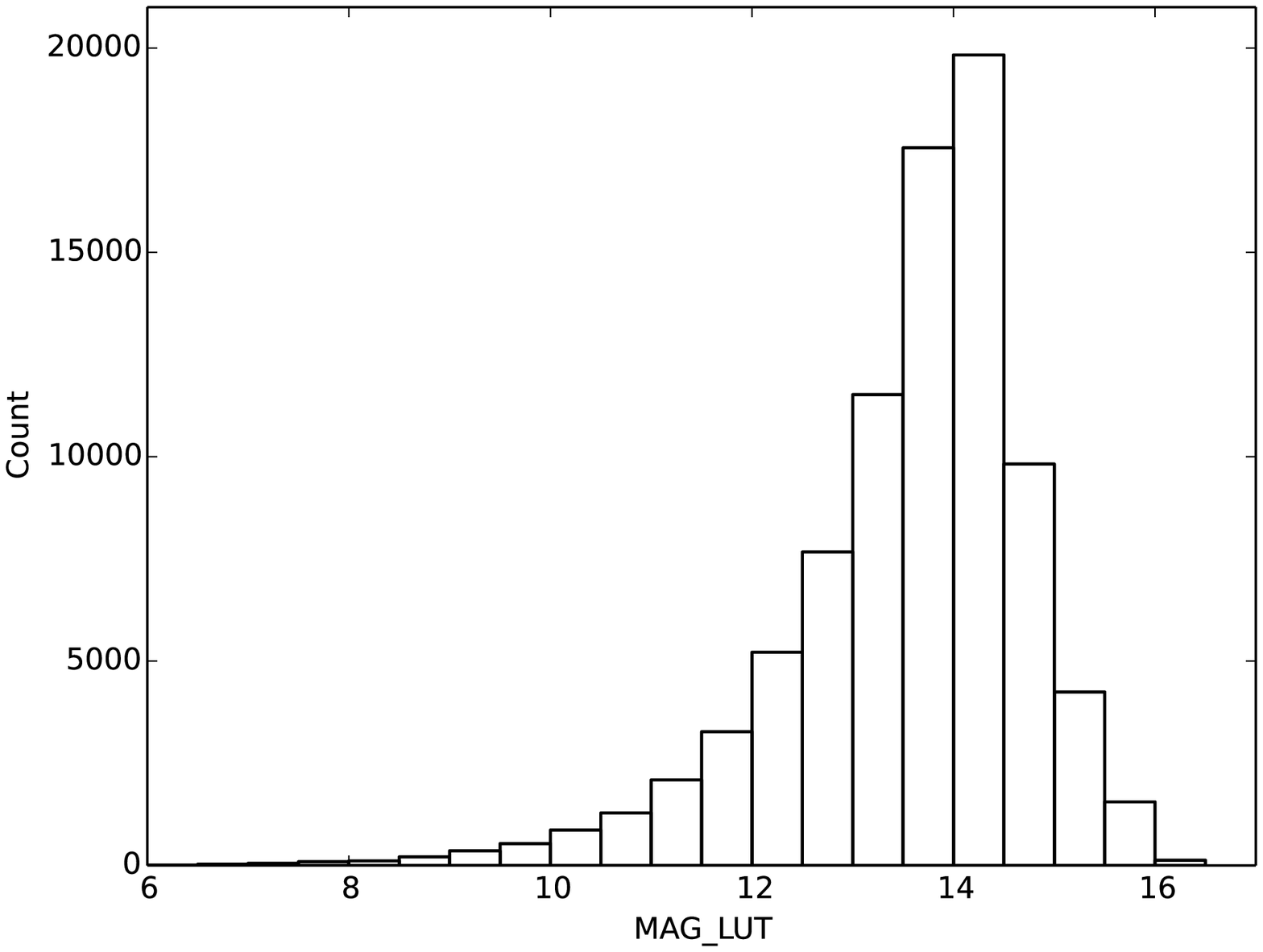}\\
  \includegraphics[width=0.6\columnwidth]{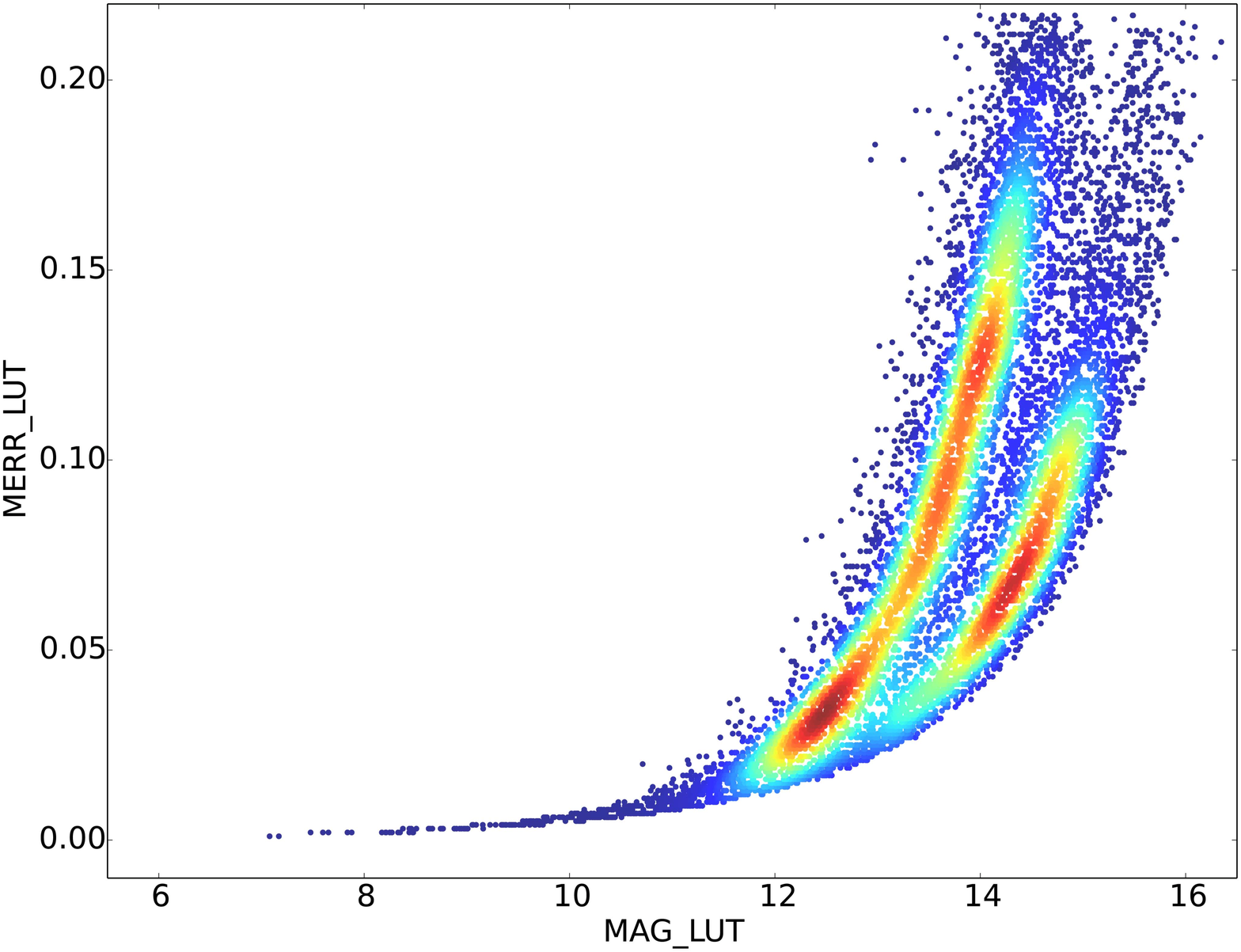}\\
  \includegraphics[width=0.6\columnwidth]{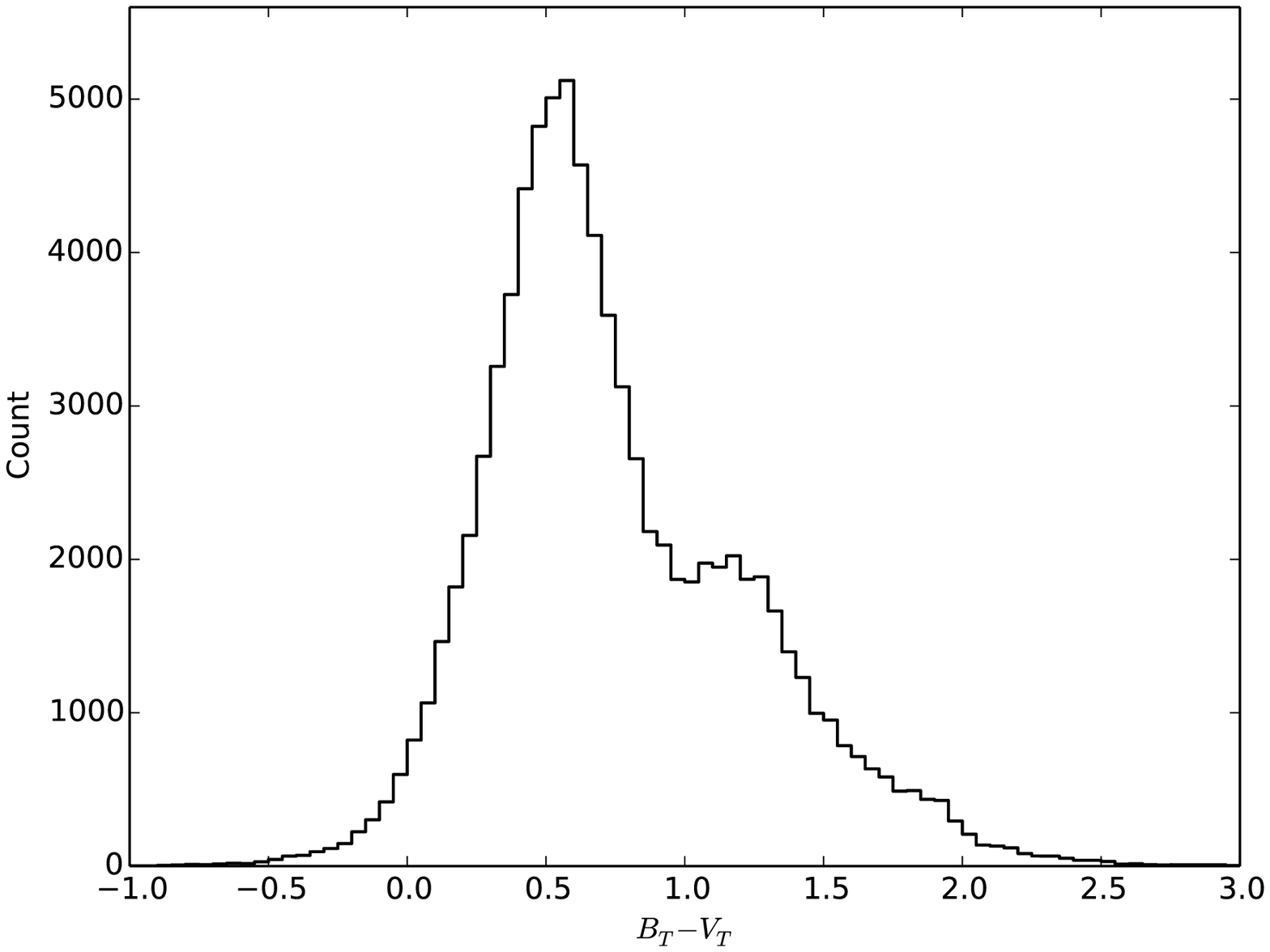}
  \caption{Some statistical properties of the catalogue.
  Top: Magnitude distribution of the stars in catalogue; 
  middle: the relation between the brightnesses and the photometric accuracies,
  the red color marks the densest region;
  there are two tracks of the relation because the right track was obtained 
  with stray light contamination much smaller;
  bottom: distribution of stars' spectral types.}
  \label{fig:statis}
\end{figure}

Some statistical properties of the catalogue are shown in Figure~\ref{fig:statis}.
The top panel of Figure~\ref{fig:statis} shows the histogram of the LUT band magnitude.
The majority of the sources in the catalogue have brightness of 13--15\,mag at the NUV band.
The faint end extends down to $\sim$16\,mag, whose corresponding sources were observed at the early
times of each lunar daytime when the stray light was at the lowest level.

The typical photometric accuracies can be derived from the middle panel of 
Figure~\ref{fig:statis}.
The diagram provides the relation between the aperture photometry uncertainties and brightness.
The plot gives a typical uncertainty of 0.2\,mag at magnitude $\sim$14.4, 
so we define 14.4\,mag as the 5$\sigma$ detection limit of the LUT survey program.

The distribution of the spectral types can be found in the bottom panel of Figure~\ref{fig:statis}.
The peak of the distribution occurs at $B_T-V_T=$~0.5--0.6\,mag, 
where F and G type stars may dominate.
A minor peak can be found around $B_T-V_T=1.2$\,mag, which K and M type stars may have.

\section{Discussion}
\label{sec:Discuss}

\subsection{Extinction}
\label{subsec:extinct}
The photometric results in the catalogue are not corrected by Galactic extinction.
The reasons are: (1) the sources are in principle stars in the Milky Way 
and extinction correction for Galactic sources is complicated. 
At least, we need to have the information of their distances, but actually they are lacking;
(2) the uncertainty of the correction should be significant.
First, even if we had their distance information, the dust/gas amounts they 
had went through are hard to estimate; 
second, for example, a certain method of galactic dust reddening estimation may 
have an uncertainty of about 16\% \citep{Schlegel1998},
which may be much larger than the uncertainty of the photometry results;
(3) the extinction correction may induce errors to photometric results, 
and these are systematic errors which come from the correction factors 
and should be treated differently from random errors.

\subsection{Aperture Correction}
\label{subsec:Apertcor}

\begin{figure}[!htbp]
  \centering
  \includegraphics[width=0.8\columnwidth]{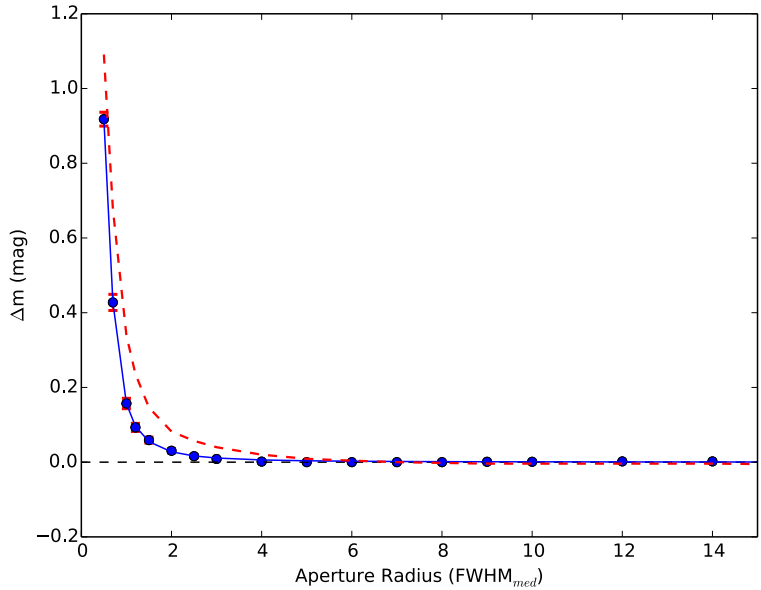}
  \caption{The curve of growth and its dispersions of LUT combined images of the survey data 
  (blue circles and red error bars). 
  The blue solid line is a Voigt model fitting to the curve of growth. 
  The red dashed line representing the curve of growth of LUT single-exposure images, 
  as has described by \citet{Meng2015ApSS}, is shown here for comparison.}
  \label{fig:surveygrowth}
\end{figure}

The catalogue doesn't include aperture corrections for photometry results.
We give the correction factor separately in this section as a choice that is subject to the users.
The reason is as follows.
Because of the image combination, 
the form of the brightness profile of stars may vary in a considerable range,
depending on the precision of the image alignment, 
profiles variation between single-exposures, field distortion, etc.
Therefore, the aperture effect of the survey data photometry is a complicated problem.
The aperture correction method for survey data follows the 
pointing observation \citep[see][Sect. 4]{Meng2015ApSS}.
We measure the ``curve of growth'' for 26 bright stars on different survey images 
and their physical positions were varying.
The scatter of the growth curves of survey program 
is obviously larger than that of pointing program,
and their profiles are different (see the median stacked profiles shown in 
Figure~\ref{fig:surveygrowth}).
Therefore, the uncertainty of aperture correction factor for survey program 
is larger than that for pointing program. 

Derived from the growth curve, the aperture correction factor for 2$\times$FWHM aperture radius
is $\Delta m_{r_{\rm 2}}=-0.029\pm0.011$,
if the aperture correction is performed with
\begin{equation}
 m_{{r_{\rm 2}},{\rm cor}} = m_{r_{\rm 2}} + \Delta m_{r_{\rm 2}}
\end{equation}
where $m_{r_{\rm 2}}$ and $m_{{r_{\rm 2}},{\rm cor}}$ are the magnitudes before and after aperture correction, respectively.

\section{Summary}
\label{sec:summary}

A star catalogue obtained from the observation data of LUT survey program is presented here,
which has 86,467 entries of stars at NUV band.
The sky coverage of the catalogue is about 2400\,$\deg^2$, 
a circular belt around the Moon's north pole, 
and a part of it has low Galactic latitude of $b<15\degr$.
An automatic pipeline is developed to process the data,
coping with stray light contamination and thereby false sources, cosmic rays, flat field calibrations, photometries, etc. 
In this first released version, 
the catalogue provides high confidence sources that have been cross-identified 
with Tycho-2 catalogue.
The SNR is constrained to be $\geq 5$, 
and the corresponding detection limit of the LUT survey is about 14.4\,mag.
Some statistical properties are given here. 
The full catalogue in electronic form is available as catalog I/335 from CDS.

\begin{acknowledgements}
This project is supported by the Key Research Program of Chinese Academy of Science (KGED-EW-603),
the National Basic Research Program of China (973-program, Grant No. 2014CB845800),
and the National Natural Science Foundation of China 
(Grant Nos. 11203033, 11473036, U1231115, U1431108). 
This project made use of 
{\sc SExtractor}, a powerful program for astronomical data analysis;
{\sc astropy}, a community-developed core {\sc Python} package for astronomy (Astropy Collaboration, 2013);
{\sc Matplotlib}, a 2D graphics package for {\sc Python} \citep{Hunter:2007};
{\sc PyRAF} and {\sc PyFITS}, products of the Space Telescope Science Institute, 
which is operated by AURA for NASA;
the SIMBAD database, operated at CDS, Strasbourg, France.
This project makes a lot of use of the tabular data analysis and visualization software 
{\sc TOPCAT} \citep{Topcat2005}
to do archiving works, which has not been described in the text.
\end{acknowledgements}

\bibliographystyle{raa}
\bibliography{msRAA-2016-0054-R1}

\end{document}